\documentstyle[epsf,twocolumn,aps]{revtex}
\def\chib{\Pi}
\def\chip{\chi_{\rm{pair}}}
\def\chis{\chi_{\rm{spin}}}
\def\TKTB{T_{KTB}}
\begin{document}
\draft
\addtolength\textheight{41pt}
\addtolength\footskip{-10pt}
\def\authornote#1{$\bullet$ {\sc #1}}
 
\twocolumn[\hsize\textwidth\columnwidth\hsize\csname @twocolumnfalse\endcsname
 
\title{Vortex-pair unbinding in the normal state of two-dimensional,
short coherence-length superconductors}
 
\author{Jan R.~Engelbrecht and Alexander Nazarenko}
\address{Department of Physics, Boston College, Chestnut Hill, MA 02167}
\maketitle
 
\begin{abstract}
We study the superconducting transition in the attractive Hubbard model
in two dimensions using the self-consistent $T$-matrix approximation.
We demonstrate that for large system sizes, this approximate
method produces $XY$ critical scaling 
in the correlation length and pair susceptibility.
For the parameters we investigate,
the critical regime is quite large, extending beyond 5 times $\TKTB$.
We discuss the role of vortex-pair unbinding in the normal state
in the context of pseudogap behaviour and the relevance to
the physics of underdoped cuprates.

\end{abstract}
 
\centerline{{\it (Submitted 19 May 1998)}}

 
]
\narrowtext

Compared to conventional superconductors, the ratio of the superconducting
pair wave-function size to the average inter-carrier spacing is many orders of
magnitude lower in the cuprates.  Consequently, the pairing in the cuprates is
much less mean-field like and quite some time ago it was argued\cite{SRE93}
that, in the cuprates, strong pairing correlations should modify the normal
state behaviour between the transition temperature $T_c$ and the mean-field
transition temperature $T^{*}$.  Early evidence for this pre-cursor pairing
was the comparison between numerical simulations\cite{NEGU_NS}
of the attractive Hubbard model with the suppression of the Knight shift and
nuclear relaxation rate\cite{NMR} between $T_c$ and room temperature in
underdoped YBCO.  The body of experimental evidence for pre-cursor pairing in
underdoped cuprates has grown considerably, for instance, optical
conductivity\cite{OPTICS} and specific heat\cite{SP-HEAT} experiments 
also indicate a loss of weight for normal state carriers between
$T_c$ and $T^{*}$.  
These measurements, involve thermodynamics or two-particle response and
are insensitive to the symmetry of the pairing.
The recent angle-resolved photo-emission (ARPES) 
results\cite{STANFORD,ARGONNE} 
have now confirmed this picture of a loss of single-particle spectral
weight in the normal state, and in addition, have confirmed a
symmetry consistent with $d_{x^2-y^2}$-pairing.

Staring from the notion that improved understanding of the cuprates 
requires theories of much stronger pairing than we are used to,
one is naturally lead to generalizations of BCS theory.
Arguably the simplest candidate for such theories is the attractive Hubbard 
model which incorporates increased fluctuations due to reduced dimensionality
with strong (albeit $s$-wave) pairing and the possibility of 
small
pair sizes.  Studies of this model have generally
focussed on either the phase diagram or the unusual
normal state.
Moreo and Scalapino\cite{Moreo_Scalapino}
followed on earlier work by Scalettar et al \cite{Scalettar_etal}
to estimate the ordering temperature by performing
finite-size scaling on the pair susceptibility.
Their Monte Carlo (MC) simulations on systems up to size $10\times10$
gave $\TKTB\approx0.045$ for $U=-4$ at quarter-filling (density $n=0.5$,
energy in units of $t$).
Subsequently, Assad et al \cite{Assad} investigated the onset of a non-zero
superfluid stiffness $\rho_s$ using MC simulations on systems of
size $8\times8$.
Very recently, Deisz et al \cite{FLEX-transition} 
used the fluctuation-exchange approximation\cite{FLEX}
(FEA) to study the superfluid stiffness and the entropy.
Most of their data was on significantly larger $32\times32$
systems and their FEA results agree well with the MC data.
However, they observe a broad onset of $\rho_s(T)$
(that does not scale as expected with system size)
instead of the discontinuity of the $KTB$ 
transition.\cite{Kosterlitz_Thouless}
The dilemma is that without a better understanding of the correlation
length it is difficult to disentangle finite-size effects,
and spurious long-range order from the expected algebraic order.

The other class of studies, originated with the
MC calculations of Randeria et al\cite{NEGU_NS}
of the temperature dependence of the spin susceptibility
$\chis(T)$, the nuclear relaxation rate $T_1(T)$ and the density of states
at the Fermi surface $N_0(T)$.
Their results agreed qualitatively with experiments on underdoped cuprates
and at an early stage suggested that precursor pairing is
manifested in the normal state of these materials.
Further investigations in this vein included the MC simulations of
Singer et al\cite{Singer} who calculated the frequency dependence
of the density of states $N(\omega)$.
In addition to 
the potentially exact MC calculations 
our earlier gaussian fluctuation studies\cite{SRE93,ERS97} were extended to
the self-consistent $T$-matrix approximation\cite{TMATRIX} (STA)
which although approximate, allows for the study of different hamiltonians
and larger systems.
From a diagrammatic point of view, the STA includes a subset of
the diagrams in the FEA.
Using this self-consistent $T$-matrix approximation we recently investigated
\cite{ENRD}
pre-cursor pairing in a model with $d$-wave symmetry that led to very
interesting predictions tested in recent ARPES experiments.\cite{NEWNATURE}

Our goal in this paper
is to study the critical behaviour of the attractive Hubbard model
in the parameter range relevant for the cuprates,
and investigate any connections between the 
Kosterlitz-Thouless-Berezinskii (KTB)
transition and the anomalous normal state properties.
We demonstrate that the self-consistent $T$-matrix approximation
is sufficiently accurate to produce
$XY$ critical scaling in the correlation length and pair susceptibility.
The model has a large critical regime 
with $\xi(T)$ falling off exponentially from $\TKTB$ until $\xi\sim 1$ 
(in units of the lattice spacing).
This happens somewhat before $T^{*}$ and above this temperature the
pre-cursor pairs are essentially uncorrelated.
Given that the STA includes a subset of the FEA diagrams
we conclude that both these approximations
capture the delicate interplay
\cite{dressed} 
between the carriers, the
precursor pairs and the unpairing of vortices.

The attractive Hubbard model hamiltonian is
\begin{equation}
H=-t\sum_{\langle ij\rangle,\sigma} 
(c^\dagger_{i\sigma}c_{j\sigma}+hc)
-|U|\sum_in_{i\uparrow}n_{i\downarrow}
\end{equation}
where the first sum runs over pairs of nearest neighbour sites on
a two-dimensional (2D) square lattice.
By varying the attraction $U$ one can study the crossover from 
BCS superconductivity in weak coupling, to the Bose condensation
of local pairs in strong coupling.
In 2D, the broken $U(1)$ symmetry belongs to the 
$XY$ universality class and one expects 
a KTB-type transition at non-zero
temperature.
On the other hand,
at half-filling,
superconductivity becomes degenerate with a charge-density-wave
instability, and this larger $SU(2)$ symmetry requires the
ordering temperature to vanish.

We study this model by self-consistently solving for the
dressed propagator and four-point vertex.  The Green's function is 
defined in terms of the self-energy
$G({\bf k},ik_n)=
\left[ 
	ik_n-\epsilon(k)+\mu-\Sigma({\bf k},ik_n) 
\right]^{-1}$.
The exact self-energy is formally defined through the
coupled equations 
\begin{equation}
\Sigma(k)
=
U\sum G(q\!\!-\!\!k)G(p)G(q\!\!-\!\!p)\Gamma_{p,k}(q)
\label{eq:sigma_form}
\end{equation}
and
\begin{equation}
\Gamma_{k,k^\prime}(q)
=
I_{k,k^\prime}(q)
-
\sum I_{k,p}(q)
G(p)
G(q\!\!-\!\!p)
\Gamma_{p,k^\prime}(q)
\label{eq:gamma_form}
\end{equation}
where the sum (with proper factors) 
represents a trace over intermediate momenta, frequencies
and, where appropriate, spins.
Here $I_{k,k^\prime}(q)$ is the irreducible particle-particle vertex,
$k=({\bf k},ik_n)$ and $q=({\bf q},iq_n)$
in a temperature formalism with Fermi and Bose Matsubara frequencies
$k_n=(2n+1)\pi T$ and $q_n=2n\pi T$ respectively.
Determining the full irreducible particle-particle vertex is 
a formidable task and we
approximate $I_{k,k^\prime}(q)$ by the bare Hubbard potential. 
Our description then reduces to the self-consistent $T$-matrix approximation.
Now $\Gamma_{k,k^\prime}(q)=\Gamma(q)$ and (\ref{eq:gamma_form}) 
simplifies to the
separable Bethe-Salpeter equation with solution
\begin{equation}
\Gamma(q)={U^2\chib(q)\over1+U\chib(q)}
\label{eq:gamma}
\end{equation}
and (\ref{eq:sigma_form}) reduces to
\begin{equation}
\Sigma(k)=\sum_{q}G(q-k)\Gamma(q)
.
\label{eq:sigma}
\end{equation}
We have subtracted the Hartree shift and
the particle-particle bubble
$
\chib(q)=\sum_{k}G(q-k)G(k)
$
is defined in terms of dressed propagators.

This amounts to keeping the subset of particle-particle diagrams in the
fluctuation exchange approximation.  The self-energy in the FEA 
includes diagrams in three channels
\par 
\vspace{5pt}
  {\epsfxsize=3.30in\epsfbox{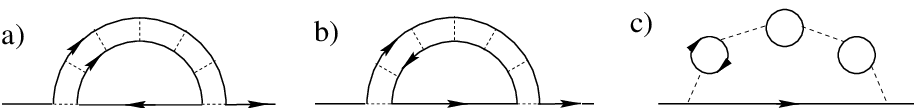}}
\par 
\vspace{5pt}
\noindent
while we only keep the particle-particle channel represented by the first term.
Our results confirm that this first class of diagrams are relevant and
the particle-particle contributions are sufficient\cite{dressed}
to produce the correct critical behaviour for the densities we
investigate.
At half-filling neither the STA or the FEA methods preserve
the degeneracy between pairing and CDW channels
that is required to restore the full $SU(2)$ symmetry.
However,
our calculations are sufficiently far from half-filling, 
such that this is not an issue.
In fact, our STA results agree quite well with
the Monte Carlo simulations\cite{NEGU_NS},
yielding a suppression of the density of states
at the Fermi level 
that tracks the reduction of spin susceptibility
similar to the Knight shift experiments.
A more detailed comparison with MC will be presented elsewhere.\cite{NERT}

Our goal here is to elucidate the nature of the normal state
pairing correlations.
Specifically we wish to investigate the
equal-time pair correlation function
\begin{equation}
\Phi({\bf r})=\langle\bar\Delta(0,0)\Delta({\bf r},0)\rangle
.
\end{equation}
This is the Fourier transform of the generalized pair
susceptibility, which in our approximation has the simple form
$
\chip(q)= \chib(q)/ [1+U\chib(q)]
$.

\begin{figure}[t]
  \vspace{-10pt}
  {\epsfxsize=3.30in\epsfbox{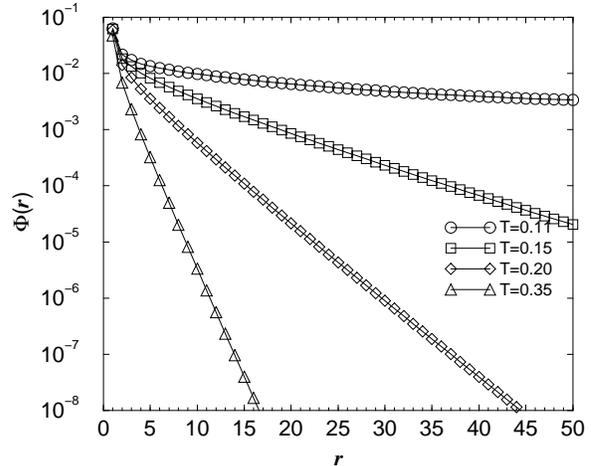}}
  \caption{
        Equal-time pair correlation function 
	$\Phi(r)$ {\it vs.}
	position $r=|(x,0)|$ in units of the lattice spacing for 
	$U=-4$, at quarter-filling, and temperatures
	$T=0.35$, 0.20, 0.15 and 0.11.
        \label{fig:xi_r}}
\end{figure}

\begin{figure}[t]
  \vspace{-10pt}
  {\epsfxsize=3.30in\epsfbox{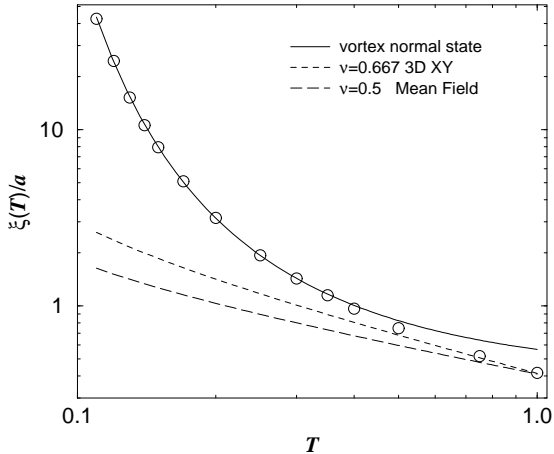}}
  \caption{Pair correlation length $\xi(T)$ {\it vs.} temperature $T$
	for $U=-4$, at quarter-filling.  The solid line
	is a fit for the exponential $XY$ scaling form. 
	The broken and dash lines represent power-law scaling
	for the mean-field and 3D $XY$ exponents 
	$\nu=\frac12$ and $\frac23$.
        \label{fig:xi_scale}}
\end{figure}

To evaluate $\chip(q)$,
we employ the standard STA/FEA procedure of 
iteratively solving for (\ref{eq:gamma}) and (\ref{eq:sigma})
on an $L\times L$ lattice using fast Fourier transforms.
We typically keep 256 Matsubara frequencies and 
to avoid finite-size effects study systems of size $L=128$.
Since this procedure uses periodic boundary conditions in space,
$\Phi(r)$ only exhibits bulk character up to distance $r\sim L/2$.
In Figure~\ref{fig:xi_r} we show our results for $\Phi(r)$ 
at four temperatures for $U=-4$ at quarter-filling.
We use these parameters for comparison with the MC data \cite{NEGU_NS}
which show a reduction of the spin susceptibility below $T=0.5$.
The correlation function indeed decays exponentially
and at large distances shows a very rapid increase upon cooling.
The correlation length $\xi$ at each temperature is defined
through the scaling relation \cite{Kosterlitz_Thouless}
\begin{equation}
\Phi(r)\sim {1\over r^\eta} \times e^{-r/\xi}.
\end{equation}
at large distances.  
To extract $\xi(T)$, the power law correction is small and
can be neglected.
(For the $XY$ universality class the exponent has the theoretical
value $\eta=0.25$ which indeed is verified below).
The temperature-dependence of $\xi(T)$ for the same attraction and density
is given as the circles in Figure~\ref{fig:xi_scale}.
The solid line is a fit to the $XY$ scaling form\cite{Kosterlitz_Thouless}
\begin{equation}
\xi(T)\sim\exp{\left(A/\sqrt{\beta_c-\beta}\right)}
\end{equation}
with $\beta=1/T$ the inverse temperature and
$\beta_c=1/\TKTB$ the inverse of the KTB transition temperature.
The best fit yields $\TKTB=0.049(2)$ for $U=-4$ and $n=0.5$.
For comparison we also show the power scaling law forms 
$\xi\sim (T-T_c)^{-\nu}$ for mean-field and 3D $XY$ exponents
$\nu=\frac12$ and $\nu=\frac23$ respectively.

The correlation length clearly scales according the the $XY$ critical 
scaling form
and upon heating decreases much faster than any power law.
It is clearly natural to interpret this scaling as due to
{\it vortex-pair unbinding} in normal state.

Furthermore, the numeric data shows critical scaling over a very
large critical regime, from $\TKTB\approx0.05$ up to $T\approx0.30$.
In fact, we observe critical scaling
until the correlation length has decayed all the way down to a 
single lattice spacing!
The MC simulations and our STA results show the suppression
of the spin susceptibility starts to develop below $T\approx0.5$.
Since $\xi$ does not exceed one lattice spacing
in the temperature range $0.35<T<0.5$ we argue that to describe
the onset of the pseudogap behaviour one needs to incorporate
amplitude fluctuations in addition to phase fluctuations.
The size of the critical regime is maximal (in the sense that
it extends down to $\xi=1$ in units of the lattice spacing) and the
very large scaling regime is reminiscent of the quantum critical 
point.\cite{SCALING}

\begin{figure}
  \vspace{-10pt}
  {\epsfxsize=3.30in\epsfbox{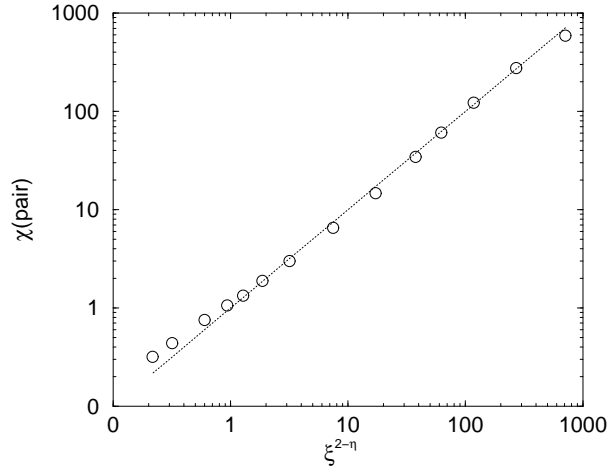}}
  \caption{Pair susceptibility $\chip(T)$ {\it vs.} power-law scaling of
	the pair correlation length $\xi(T)^{2-\eta}$ for $\eta={1\over4}$.
        \label{fig:chip_scale}}
\end{figure}

A further test for $XY$ scaling is to consider the uniform, static, 
pair susceptibility which,
in our approximation,
is proportional to $\Gamma(q)$ with $q=0$ in Eq.~\ref{eq:gamma}.
Thus $\chip$ diverges
as $\lambda=-U\Pi(0)\to1$.  While $\lambda$ reaches the value one
at a non-zero temperature in mean-field theory, 
more accurate calculations such as MC, FEA or STA
should never give $\lambda=1$ for a finite system.  Indeed, our calculations
agree with the FEA calculation of Ref.\cite{FLEX-transition} yielding 
$\lambda<1$ for {\it all} temperatures.  
In fact, this even holds in three dimensions clearly due to
the absence of phase transitions in finite systems.
However, provided we keep the system size $L>\xi(T)$, we can observe the 
approach to the divergence in $\chip$ as $\TKTB$ is approached from above.
In Figure~\ref{fig:chip_scale} we plot 
$\chip(T)$ {\it vs} the temperature-dependent correlation length
$\xi(T)$ to the power $2-\eta$ for $U=-4$ and $n=0.5$ at several temperatures.

We have explicitly set $\eta=\frac14$ -- the universal $XY$ value.  A
fit to the lower temperature data gave $\eta\approx0.248$.  
The error in $\chip$ is larger due to the
denominator of (\ref{eq:gamma}) as $\lambda\approx1$ at low temperatures.
None the less, our data shows $XY$ critical scaling over nearly three orders 
of magnitude.

The degree to which both $\xi(T)$ and $\chip$ displays
$XY$ scaling is remarkable.  
This demonstrates that the self-consistent $T$-matrix approximation
which includes self-energy corrections 
but neglects vertex corrections still captures\cite{dressed}
the `relevant' fluctuations.
Further, the very large critical regime in the attractive Hubbard Model
vindicates our assertion of the importance of small distance physics 
(which is neglected in low-energy, long-wavelength expansions).

We have been maintaining for some time\cite{SRE93,NEGU_NS,LANL_conf} that as a 
model of
a 2D short coherence length superconductors, the unusual normal state
properties of the attractive Hubbard Model is of relevance to the
high-$T_c$ superconductors.  This is especially true for the 
underdoped cuprates which are much more\cite{ANISOTROPY}
two-dimensional than their optimally-doped or overdoped counterparts.

It is therefore natural to conclude that our results have
far-reaching implications for the cuprates.
The picture that emerges is that upon cooling the normal state
behaviour is strongly influenced by the 2D $XY$ critical point
over a large temperature range,
until three-dimensional coherence starts to develop just above
the observed (3D) critical temperature.
Thus, the strong 2D pairing correlations imply that 
(zero-field\cite{VORTICIES})
vortex-pair unbinding is important in the normal state of the cuprates
and that vortices provide an additional (strong) scattering mechanism.
This scattering channel disappears below $T_c$ and although this $s$-wave
model of course does not contain the $d$-wave pairing observed in
ARPES experiments, this framework is clearly consistent with the
observation of broad spectral peaks in the normal state
which sharpen dramatically below $T_c$.
The short distance over which phase-coherence is established
then also explain why very strong paraconductivity is not observed over a 
large temperature range.

These results also emphasize that in order to study the onset of
order numerically, one needs very large systems to eliminate 
finite size effects.  
The transition observed by Deisz et al\cite{FLEX-transition} is therefore
indeed a KTB type transition although the $32\times32$ systems they
focussed on may be on the borderline of being influenced
by finite-size effects.
Whereas the STA/FEA methods
may have problems near half-filling  where the model has $SU(2)$
symmetry, these methods are very successful at keeping
track of relevant pairing correlations away from half-filling
while ensuring that large correlation lengths do not mimic
spurious long-range order.
These methods should therefore be used in studying a variety of models
to complement the MC calculations which are restricted to smaller lattices, 
but of course have other advantages.

It should be noted that several other groups have emphasized 2D pair 
correlations in the cuprates, dating back to Uemura's\cite{UEMURA}
investigation of the superfluid density.  Also, the physics we discuss 
here and introduced in Refs.~\cite{SRE93,NEGU_NS,LANL_conf,ENRD} 
bears some resemblance to the framework 
developed later by Emery and Kivelson\cite{EMERY} -- although they
emphasize the proximity of an insulating state in the cuprates.
The attractive Hubbard model is, however, also susceptible to
a metal-insulator-transition near half-filling due to the
charge-density-wave correlations, although this simple model may not 
capture the doped Mott insulator physics of the cuprates.

Above we have demonstrated the connection between $XY$ fluctuations and
pseudogap physics. It is important to emphasize that it is crucial to also
be in the intermediate coupling (short coherence length) regime of the phase 
diagram.  In the weak coupling limit $XY$ fluctuations do not produce 
pseudogap behaviour over an appreciable temperature range.\cite{VARENNA}

To summarize we have demonstrated that
the self-consistent $T$-matrix approximation
can capture the normal state $XY$ physics of the unbinding of vortex pairs
for the attractive Hubbard model in 2D.
The scaling regime is very large, extending beyond 5 times $\TKTB$,
until the correlation length has decreased down to a single lattice spacing.
The large regime influenced by vortex-pair unbinding 
suggests that the excitations may be relevant to the normal state
pseudogap behaviour of underdoped cuprates.
The large window of temperatures where $\xi(T)\gg10$
also implies that when investigating the development of order,
it is important to complement MC calculations
with STA/FEA calculations on large systems.

We would like to acknowledge many interactions with Mohit Randeria
which strongly influenced the framework presented here, as well as
fruitful discussions with Elbio Dagotto and Miklos Gul\'acsi. 
A.~N. and J.~E. are supported by Boston College.




\begin{references}
\bibitem{SCALING}
It is tempting to speculate that the large scaling regime may be
related to the line of critical points extending down from
$\TKTB$ to zero temperature. The proximity of a
$T=0$ critical point at half-filling may also be important.
\bibitem{VORTICIES}
It should be emphasized that we are considering a model of a neutral
superfluid; there is no penetration depth in the model and 
we are discussing zero-field vortices.
\end{references}
\end{document}